\documentclass[PP]{AEA}
\usepackage{fullpage,amsmath,amssymb,amsfonts,xcolor,natbib}
\usepackage{algorithm,bbm,graphicx}
\usepackage[hidelinks]{hyperref}
\usepackage[]{algpseudocode}
\usepackage{booktabs,colortbl,cancel,multirow,natbib}

\usepackage{mathtools, cuted}
\def\MD{\textsc{md}}
\def\SMD{\textsc{smd}}
\def\rnr{r\textsc{nr}}
\def\rqn{r\textsc{qn}}
\def\thetab{\theta_{b}}

\def\gb{G_b}

\def\dconv{\smash{\mathop{\longrightarrow}\limits^d}}     

\def\var{\text{var}}
\newcolumntype{a}{>{\columncolor{Gray}}c}
\newcolumntype{b}{>{\columncolor{white}}c}
\newcommand{\mc}[2]{\multicolumn{#1}{c}{#2}}

\definecolor{Gray}{gray}{0.85}
\definecolor{LightCyan}{rgb}{0.88,1,1}
\def\dmk{\textsc{dmk}}

\usepackage{caption}

\draftSpacing{1.5}

\begin{document}

\title{{\bf Estimation and Inference by Stochastic Optimization:\\ Three Examples}}
\shortTitle{Estimation and Inference by Stochastic Optimization: Three Examples}
\author{Jean-Jacques Forneron\thanks{Department of Economics, Boston University, 270 Bay State Rd, MA 02215 Email: jjmf@bu.edu}  and Serena Ng\thanks{Department of Economics, Columbia University and NBER, 420 W. 118 St. MC 3308, New York, NY 10027 Email: serena.ng@columbia.edu
\newline  Financial Support from the National Science Foundation (SES 1558623, 2018368) is gratefully acknowledged. \newline The authors would like to thank Elie Tamer for useful comments and suggestions.} }

\pubMonth{Month}
\date{\today}
\pubYear{Year}
\pubVolume{Vol}
\pubIssue{Issue}
\maketitle
\captionsetup[table]{name=Table} \setcounter{table}{0}




%



Repeated  optimizations required to compute parameter estimates and bootstrap standard errors of complex models can be computationally burdensome.  In \citet{jjng-rnr}, we design a resampled Newton-Raphson  algorithm (\rnr)  that provides  consistent estimates and valid standard errors in  {\em one run} of the optimizer. The key insight is that the algorithm  serves as  a resampling device to produce a Markov chain of iterates with desirable properties.  In this paper, we illustrate that \rnr\; can speed up  BLP estimation from almost five hours using standard ($n$ out of $n$)  bootstrap to just over an hour and can be further reduced to fifteen minutes using a  resampled quasi-Newton (\rqn) algorithm that does not directly compute the Hessian. A Monte-Carlo exercise using Probit IV regressions shows that \rnr\; and \rqn\; provide accurate estimates and coverage. The appeal of the proposed approach  goes beyond faster computation. A re-sampling based indirect inference estimator not only produces standard errors easily, but is also more efficient than one obtained by classical optimization. This is illustrated by a dynamic panel model example.

\section{The Setup}

Many economic applications entail minimizing a sample objective function $Q_n(\theta)$ with respect to a vector of parameters $\theta$ to obtain an estimate $\hat \theta_n = \text{argmin}_{\theta} Q_n(\theta).$ Under regularity conditions, $\hat\theta_n$ is $\sqrt{n}$ consistent for the true value $\theta^\dagger$ and $(\mathbb V^\dagger)^{-1/2}\sqrt{n}(\hat\theta_n-\theta^\dagger)\dconv \mathcal  N(0,I_d)$.  The sandwich variance $\mathbb V^\dagger$ required for inference depends on both the gradient and the Hessian which are often analytically intractable. Bootstrap inference approximates the asymptotic distribution but requires repeated optimization  each time a  batch of data of size $n$ is resampled. Alternatives 
are available to speed up computation but they still necessitate a preliminary estimate $\hat\theta_n$.

The  Newton-Raphson algorithm computes $\hat\theta_n$  by iterating until convergence:  \[\theta_{k+1}=\theta_k - \gamma_k [H_n(\theta_k)]^{-1} G_n(\theta_k),\]  where $\gamma_k$ is a learning rate, $G_n(\theta_k)$ is the gradient,  and the conditioning matrix is set to  the inverse of the Hessian  $H_n(\theta_k)$ so that $[H_n(\theta_k)]^{-1}G_n(\theta_k)$ determines the direction of the update. In \citet{jjng-rnr}, we propose a novel resampled  Newton-Raphson algorithm (\rnr)   that  produces  an estimate of $\theta$ and its standard errors in one run of the optimizer. 
\paragraph{Algorithm \rnr}
\begin{itemize}
\item[1.] {\bf Inputs:} (a)  initial guess $\theta_{0}$; (b)  bootstrap sample size $B$ and burn-in period \textsc{burn}; (c)  batch size $m\leq n$, and (d)
 fixed learning rate $\gamma \in (0,1]$.        
\item[2.] {\bf Resample:}  For $b=1,\dots,\textsc{burn}+B$
\begin{itemize}
 \item[a.]  Resample a $(b+1)$-th batch of data of size $m$,
 \item[b.]  Update  $H_b=H_m^{b+1}(\thetab)$ and $G_b= G_m^{(b+1)}(\thetab)$,
 \item[c.] Update $\theta_{b+1} = \thetab - \gamma H_b^{-1} \gb.$
\end{itemize}
\item[3.] {\bf Outputs:} Discard the first \textsc{burn} draws. Let $\phi(\gamma)=\frac{\gamma^2}{1-(1-\gamma)^2}$ and output
\begin{itemize}
\item[a.] $\overline{\theta}_{\rnr}=\frac{1}{B} \sum_{b = 1}^B \thetab$,
\item[b.] $V_{\rnr}=\frac{m}{\phi(\gamma)}\widehat{\var}(\theta_b)$ where    $\widehat{\var}(\theta_b) =\frac{1}{B} \sum_{b = 1}^B (\thetab-\overline{\theta}_{\rnr})(\thetab-\overline{\theta}_{\rnr})^\prime.$
\end{itemize}
\end{itemize}
The main idea of \rnr\ is to combine estimation with inference by exploiting the randomness due to re-sampling \textit{within} the optimizer. Each  $b+1$-th sample consists of  $m$ observations drawn randomly from the original data.\footnote{For instance, iid data can be re-sampled at the individual level with replacement. Clustered data, as in Example 1 below, should be re-sampled at the cluster level with replacement.} Then $Q_m^{(b+1)}(\theta)$ is evaluated,  its  gradient  $G_m^{(b+1)}(\theta_b)$  and Hessian $H_m^{b+1}(\thetab)$  are  used to update $\theta_b$ to $\theta_{b+1}$.
The estimator is computed by taking the mean over draws after discarding the first $\textsc{burn}$  iterates to reduce the impact of the initial guess $\theta_0$.   Standard errors are obtained from  the draws after   a sample size and scale adjustment of $\sqrt{\frac{m}{\phi(\gamma)}}$.  Like MCMC, inference is sampling-based  but the approach is fully frequentist. Unlike other bootstrap shortcuts, our approach does not require a preliminary estimate $\hat\theta_n$.\footnote{See e.g. \citet{Davidson1999}, \citet{Andrews2002}, \citet{Kline2012}, \citet{Honore2017}.}

Evaluating the direction of change using small batches of data  is in the spirit of stochastic optimization, but there are two important differences. First, while the learning rate $\gamma_b$ in stochastic optimization declines with each $b$,  our $\gamma\in(0,1]$ is constant. This allows us to analytically establish that the draws  $\{\theta_b\}_{b=1}^B$  form a Markov chain with stationary ergodic properties. Second,  whereas  stochastic optimization uses $m$ fixed (as small as one) with  efficient  computation as a goal, we also have inference in mind which necessitate $m$ to increase faster than $\sqrt{n}$. Statistical and computational efficiency are conflicting goals in this context.

Under certain conditions in \citet{jjng-rnr},  the \rnr\; draws   have  two properties:
\begin{align*} \arraycolsep=1.0pt
      \begin{array}{rlr}
      \sqrt{n}\left( \overline{\theta}_{\rnr} - \hat\theta_n  \right) &= o_{p^\star}(1), &\text{(Estimation)}\\
      V_{\rnr}^{-1/2}\sqrt{m}\left( \theta_b - \hat\theta_n \right) 
      &\overset{d^\star}{\to} \mathcal{N}\left( 0, I_d \right), &\text{(Inference)}
      \end{array}
\end{align*}
where $V_{\rnr}$ is defined in Algorithm  above.  The first (consistency) result states that the mean estimator $\bar \theta_{\rnr}$ is first order equivalent to the classical estimator  $\hat\theta_n$.   The second (inference) result  states that the distribution of the draws is  first-order equivalent to that of $\theta^{(b)}_m$, the bootstrap distribution. A sketch of the argument is as follows. For the same resampling scheme, it is known that the standard bootstrap yields valid inference. We  show  that when   $m$ and $\gamma \in (0,1]$ are appropriately chosen,  the distribution of the \rnr\; draws  is close to that  of the standard bootstrap up to scale, and by implication, close to the limiting distribution of $\hat\theta_n$. But unlike the standard bootstrap which needs repeated optimizations, \rnr\; produces standard errors in the same  optimization that  produces estimates $\bar\theta_{\rnr}$.  
This means that upon completion of that single run, a $(1-\alpha)\%$ confidence interval for the $j$-th coefficient can be immediately constructed as:  \[(\overline{\theta}_{\rnr,j} + q_{\alpha/2},\overline{\theta}_{\rnr,j} + q_{1-\alpha/2})\]
where  $q_{\alpha/2}$ is the  $\alpha/2$ quantile of  $\sqrt{\frac{m}{n\phi(\gamma)}}(\theta_{b,j} - \overline{\theta}_{\rnr,j})$. Wald statistics can also be computed using $V_{\rnr}$ as a plug in estimate of $\mathbb  V^\dagger$.

The two results above also hold for a faster resampled quasi-Newton algorithm, called rQN, which approximates the Hessian by a least-squares interpolation scheme, it is described in Forneron and Ng (2020). This scheme ensures the conditioning matrix is both symmetric and positive definite which is required for inference.  Though the consistency result  also holds for many conditioning matrices, the  inferential  result  only holds for  conditioning matrices that  approximate the inverse Hessian sufficiently well because the sandwich variance structure cannot be replicated otherwise. Thus, resampled gradient descent which uses an identity matrix for conditioning will give incorrect standard errors but valid estimates.

Algorithms \rnr\ and \rqn\ are especially useful when the  model is costly  to optimize. But they also have  statistical appeals:-  the draws are immediately available for post estimation diagnostics,  and in the case of simulation estimation, $\bar\theta_{\rnr}$  can even be  more efficient than an estimate obtained from classical optimization. We now illustrate some of these properties.

\begin{table}[ht] \caption{Demand for Cereal: Estimates and Standard Errors (Random Coefficients)} \label{tab:blp}
      \centering
      \begin{tabular}{cl|baa|bbaa}
        \hline \hline 
        & & \multicolumn{3}{c|}{Estimates} & \multicolumn{4}{c}{Standard Errors} \\ \hline
        & & $\hat\theta_n$ & \mc{1}{\rnr} & \multicolumn{1}{c|}{r\textsc{qn}} & \textsc{boot} & \dmk &  \mc{1}{\rnr} & \mc{1}{r\textsc{qn}} \\ 
        \hline \parbox[t]{2mm}{\multirow{4}{*}{\rotatebox[origin=c]{90}{stdev}}}
        & const. & 0.284 & 0.263 & 0.273 & 0.129 & 0.127  & 0.123 & 0.120 \\ 
        & price  & 2.032  & 2.188 & 1.983 & 1.198 & 1.026  & 0.975 & 0.950 \\ 
        & sugar & -0.008 & -0.006 & 0.006 & 0.017 & 0.012 & 0.012 & 0.012 \\ 
        & mushy & -0.077 & -0.055 & -0.044 & 0.177 & 0.168 & 0.166 & 0.167 \\ \hline \parbox[t]{2mm}{\multirow{4}{*}{\rotatebox[origin=c]{90}{income}}}
        & const. & 3.581 & 3.464 & 3.646 & 0.666 & 0.738 & 0.714 & 0.662 \\ 
        & price & 0.467 & 1.335 & 0.111 & 3.829 & 4.275 & 4.040 & 3.569 \\ 
        & sugar & -0.172 & -0.171 & -0.174 & 0.028 & 0.028 & 0.027 & 0.031 \\ 
        & mushy & 0.690 & 0.647 & 0.694 & 0.345 & 0.346 & 0.339 & 0.333\\  \hline
        \multicolumn{2}{c|}{} & \mc{1}{} & \mc{1}{} & \multicolumn{1}{c|}{time} & 4h36m & 1h1m & 58m & 15m \\      
         \hline \hline
      \end{tabular}
\end{table}

\section{Example 1:  Demand for Cereal}

We consider the BLP model of \citet{Berry1995} for the cereal data generated in \citet{nevo:00}. The data consists of market shares $s_{gj}$ in market $g \in \{1,\dots,94\}$ for product $j \in \{1,\dots,24\}$.  Parameters on terms that enter linearly are projected out by 2SLS. We then drop interaction terms that seem difficult to identify.  This leaves us with $d=8$ parameters that enter the moment conditions $g$ non-linearly. Evaluation of the objective and its gradient is costly because  fixed-point iterations are needed to invert market shares.\footnote{We use the \textsc{BLPestimatoR} R package which builds on C++ functions to evaluate the GMM objective and analytical gradient \citep{brunner2017}.}   
We perform $m$ out of $n$ resampling  at the market level. This level of clustering controls for possible correlations in the unobservables at the market level.  That is,  for each $b=1,\dots,B$ we draw markets $g^{(b)}_{1},\dots,g_{94}^{(b)}$ from $\{1,\dots,94\}$ with replacement,  taking the associated shares and characteristics $\{ s_{g^{(b)}j},X_{g^{(b)}j}\}_{j=1,\dots,24}$ as observations within each market.  
 We set $\gamma = 0.2$ and $\textsc{burn}=10$ draws. Since the number of clusers is relatively small, we set $m=n=94$.

Table \ref{tab:blp} indicates that the \rnr\, estimates  are similar to  $\hat\theta_n$ obtained from classical optimization.  The standard errors are  similar  across methods but the \rnr\; ones are  nearly 5x faster to compute than the bootstrap and are comparable to \citet{Davidson1999}, denoted as \dmk, even excluding the time used to get the preliminary estimate. The \rqn\;  further reduces computation time over \rnr\; by a factor of 4.   

The  estimates based on classical optimization reported above use only  20 integration draws as in \citet{nevo:00}. More accurate estimates will require more draws, so the gains in using \rnr\; and \rqn\; are conservative. Besides inference on the parameters, the \rnr\; draws can also be useful in post-estimation analysis. For instance, in more involved counterfactuals such as merger analyses, the delta-method can be challenging to apply while re-evaluating counterfactuals on bootstrap draws is straightforward.  

\begin{table}[ht]
      \caption{Probit IV: finite sample properties in estimation and inference} \label{tab:piv}
      \centering
      \begin{tabular}{l|aa|aa|baa}
        \hline \hline
       & \multicolumn{2}{c|}{Average Estimate} & \multicolumn{2}{c|}{Standard Deviation} & \multicolumn{3}{c}{Rejection Rates}\\
     m  & \mc{1}{\rnr} & \multicolumn{1}{c|}{\rqn} & \mc{1}{\rnr} & \multicolumn{1}{c|}{\rqn} & \mc{1}{\textsc{boot}} & \mc{1}{\rnr} & \mc{1}{\rqn} \\ 

        \hline
        \multicolumn{8}{c}{$\gamma = 0.2$}\\
        \hline
        500 & 1.033 & 1.037  & 0.211 & 0.212 &  0.068 & 0.070 & 0.067 \\
        100  & 1.022 & 1.042  & 0.218 & 0.219  & 0.083 & 0.068 & 0.070 \\ 
        50  & 1.003 & 1.072  & 0.217 & 0.459 & 0.082 & 0.042 & 0.050 \\  \hline
        \multicolumn{8}{c}{$\gamma = 0.1$}\\ \hline
        500  & 1.032 & 1.033 & 0.211 & 0.210 & - & 0.048 & 0.051 \\ 
        100 & 1.029 & 1.036  & 0.219 & 0.219  &- & 0.052 & 0.058 \\  
        50   & 1.024 & 1.043  & 0.222 & 0.223 & - & 0.055 & 0.057 \\
        \hline \hline
      \end{tabular} 
\end{table}

\section{Example 2: Probit IV Regression}
The second example uses simulations to evaluate the finite sample properties of \rnr\ and \rqn. We consider a probit instrumental variable regression model specified as
\begin{align*}
      &y_{1i} = \mathbbm{1}\{ \alpha y_{2i} + \beta_0 + \beta_1x_i + \rho v_i + u_i \},\\ &y_{2i} = \xi_0 + \xi_1x_i + \pi z_i + v_i,
\end{align*}
where $x_i,z_i$ are independent and exponentially distributed with rate $1$; $v_i,u_i$ are independent standard normal; $\theta^\dagger = (\xi_0,\xi_1,\pi,\alpha,\beta_0,\beta_1,\rho) = (0,1,1,1,0,1,1)$. These seven coefficients are jointly estimated in a just-identified GMM system using the sample vector of moments:
\[ \overline{g}_n(\theta)  = \frac{1}{n} \sum_{i=1}^n \left( \begin{array}{rcl} r_{1i}(\theta) &\otimes& (1,x_i,z_i,r_{2i})^\prime \\ r_{2i}(\theta) &\otimes & (1,x_i,z_i)^\prime \end{array} \right), \]
where $\otimes$ is the kronecker product, $r_{2i}(\theta) = y_{2i} - (\xi_0 + \xi_1x_i + \pi z_i)$ and  $r_{1i}(\theta) = y_{1i} - \Phi( \alpha y_{2i} + \beta_0 + \beta_1x_i + \rho r_{2i} )$. We set $n=500$ in each of the 1000 Monte-Carlo replications. The estimates $\hat\theta_n$ are computed using the  \textsc{bfgs} routine in \textsc{r}.  For \rnr, \rqn\, we use $B=2000$ and  consider $\gamma \in \{0.2,0.1\}$,  $m \in \{50,100,500\}$. For \textsc{bfgs}, \rnr\ and \rqn\,  $\theta_0 = (0,\dots,0)$. For the bootstrap, optimization is initialized at $\hat\theta_n$ and we only use   $B=500$ as is  common practice, but even this  is  slower than \rnr\ and \rqn\ with $B=2000$.

Table \ref{tab:piv} compares the properties of the estimates and quantile-based confidence intervals for $\alpha$, the coefficient on the endogenous regressor $y_{2i}$ which is typically a parameter of interest. The average estimates and standard errors with classical estimation $\hat\theta_n$ are $1.034$ and $0.210$ which are generally comparable to \rnr\ and \rqn\ reported in the table.  The exception is the $m=50$, $\gamma=0.2$ case which can be attributed to three replications for which  $\textsc{burn}=50$ appears to be too small. For coverage, the usual $\hat\alpha_n \pm 1.96 \cdot \text{se}(\hat\alpha_n)$ confidence interval is very close to the 95\% level with a rejection rate of $0.055$. 
Using $m=500$ and $\gamma=0.2$, the coverage of \rnr, \rqn\ is comparable to that of the bootstrap. For $m<n$ the accuracy of the bootstrap declines while \rnr\ and \rqn\ are less affected. For $\gamma=0.1$, coverage is closer to the nominal 95\% confidence level for the entire range of $m$ and $\gamma$ values. 

\section{Example 3: Simulation-based Estimation}
 The third example highlights the statistical gains of \rnr/\rqn\ for simulation-based estimation. Consider  the linear dynamic panel model:
\begin{align}
      &y_{it} = \rho y_{it-1} + \beta x_{it} + \alpha_i + \sigma e_{it}, \label{eq:dynpan}
\end{align}
where $t\in\{1,\dots,T\}, i\in\{1,\dots,n\}$, $\theta = (\rho,\beta,\sigma)$.  The least-squares dummy variable (\textsc{lsdv}) estimator is inconsistent as $n\to\infty$ with $T$ fixed.  \citet{Gourieroux2010} consider simulation  estimation of $\theta$ using $\hat\psi_n = \hat\theta_{n,\textsc{lsdv}}$ as auxiliary statistics. 
Given  draws $e_{it}^s$ and a value of $\theta$, simulate $S$ panels of $y_{it}^s$ of size $(n,T)$ using (\ref{eq:dynpan}), compute the simulated  moments $\hat\psi_n^s(\theta) = \hat\theta_{n,\textsc{lsdv}}^s$. The indirect inference (\textsc{ind}) estimator  $\hat\theta_{n,\textsc{ind}}^S = \text{argmin}_{\theta} \|\hat\psi_n- \frac{1}{S}\sum_s \hat\psi_n^s(\theta)\|$ has an automatic bias correction property and is consistent as $n\rightarrow\infty$ even if  $T$ is fixed, but its  variance is inflated by a factor $(1+\frac{1}{S})$ due to simulation noise.  The $n$ out of $n$ bootstrap is often used to obtain standard errors of indirect inference estimates. Throughout the estimation above, the covariates $x_{it}$ and the simulation draws $e_{it}^s$ are fixed while the optimizer solves for $\hat\theta^S_{n,\textsc{ind}}$.
\begin{table}[ht]

      \centering \caption{Dynamic Panel: finite sample properties in estimation and inference} \label{tab:panel}
{ 
\begin{tabular}{l|aa|aa|baa}
      \hline \hline
      & \multicolumn{2}{c|}{Average Estimate} & \multicolumn{2}{c|}{Standard Deviation} & \multicolumn{3}{c}{Rejection Rates}\\
      \hline
      m  & \mc{1}{\rnr} & \multicolumn{1}{c|}{\rqn} & \mc{1}{\rnr} & \multicolumn{1}{c|}{\rqn} & \mc{1}{\textsc{boot}} & \mc{1}{\rnr} & \mc{1}{\rqn} \\
      \hline
        \multicolumn{8}{c}{$\gamma = 0.1$, $S = 1$}\\
      \hline
      500 & 0.599 & 0.599 & 0.023 & 0.023 & 0.052 & 0.044 & 0.045 \\ 
      100 & 0.598 & 0.599 & 0.023 & 0.023 & 0.054 & 0.037 & 0.047 \\ 
      \hline
        \multicolumn{8}{c}{$\gamma = 0.1$, $S = 10$}\\
      \hline
      500 & 0.600 & 0.600 & 0.023 & 0.023 & 0.049 & 0.044 & 0.043 \\ 
      100 & 0.598 & 0.598 & 0.023 & 0.023 & 0.047 & 0.043 & 0.041 \\ 
         \hline \hline
      \end{tabular}
      }
\end{table}

In contrast, \rnr\ and \rqn\ resample $m$ out of $n$ individual paths of $(x_{it})_{t=1,\dots,T}$ and simulate new draws $e_{it}^{s,b}$ at each iteration $b$. 
This has two advantages. First, as in the examples above, it is faster than the conventional bootstrap in producing standard errors. Second,  the simulation noise across $b$ averages out, and as a consequence, \rnr/\rqn\ achieve the same asymptotic variance as an \textsc{ind} estimator that uses $S=\infty$ simulations.  This statistical efficiency gain comes for free since we only use finitely many $S$ simulated samples at each iteration $b$, and with $m$ possibly less than $n$.  The only proviso is that a second chain of draws is needed to produce correct standard errors and confidence intervals, as shown in \citet{jjng-rnr}. 

To illustrate, data are simulated with $(\rho,\beta,\sigma) = (0.6,1,1)$,  and $x_{it},e_{it}$ are iid standard normal with $n=500$ and $T=5$. We use $m \in \{100,500\}$, $B=2000$, $S \in \{1,10\}$, $\textsc{burn}=45$, and $\gamma = 0.1$ for \rnr, \rqn.  For the bootstrap we only use $B=500$ as is common practice. The \textsc{lsdv} estimate $\hat\rho_{n,\textsc{lsdv}}$ is $0.306$ on average with standard deviation $0.017$, exhibiting significant downward bias from the true $\rho = 0.6$. \textsc{ind} removes the downward bias almost entirely with an average estimate of $0.599$ and $0.601$ for $S=1,10$, respectively. The standard deviation of the \textsc{ind} estimates is $0.032$ and $0.024$ for $S=1,10$. Table \ref{tab:panel} shows that \rnr\ and \rqn\ preserve this bias correction and have smaller standard deviations even with $S=1$ and $m<n$, as predicted by theory. Coverage is close to the nominal 95\% level for the usual $\hat\theta_n \pm 1.96 \cdot \text{se}(\hat\theta_n)$ confidence interval, with rejection rates of $0.049$ and $0.046$ for $S=1,10$. Bootstrap, \rnr, and \rqn\ have similar coverage. Increasing $S$ has little effect on  \rnr, \rqn\ but improves the accuracy of \textsc{ind}. 

\bibliographystyle{aea}
\bibliography{refs}

\end{document}